# Absence of Long Range Magnetic Order in the $La_{1.4}Sr_{0.8}Ca_{0.8}Mn_2O_7$ Bilayered Manganite

*Lorenzo Malavasi[a*], Maria Cristina Mozzati[b], Vladimir Pomjakushin[c], Cristina Tealdi[a], Carlo Bruno Azzoni[b], and Giorgio Flor[a]*

[a]Dipartimento di Chimica Fisica "M. Rolla", Università di Pavia, V.le Taramelli 16, I-27100, Pavia, Italy.

[b]CNISM, Unità di Pavia and Dipartimento di Fisica "A. Volta", Università di Pavia, Via Bassi 6, I-27100, Pavia, Italy.

[c]Laboratory for Neutron Scattering, ETHZ & PSI, CH-5232, Villigen PSI, Switzerland.

In this work we studied, by means of high-resolution neutron diffraction as a function of temperature, the $La_{1.4}Sr_{0.8}Ca_{0.8}Mn_2O_7$ bilayered manganite for two different annealing treatments. Out data allowed us to shown, for the first time, the absence of long-range magnetic order in this optimally doped bilayered manganite where the A-site of the structure is doped with equal proportions of different isovalent cations (Ca and Sr). The system, however, presents defined IM transitions which suggest that the transport properties are not linked to the evolution of long-range order and that two dimensional spin ordering in the layers of the perovskite blocks may be sufficient to "assist" the hole hopping.

Possible reason for the suppression of magnetic order induced by the Ca doping is a size effect coupled to the cation size mismatch between the Sr and Ca ions.

# Introduction

There is an increasing interest in materials that show magnetoresistance, because of their use in magnetic information storage or as magnetic field sensors [1]. Most of the available theoretical and experimental work has, until now, been focused on the 3D perovskite structures, that is the $n=\infty$ end-member of the $A_{n+1}B_nO_{3n+1}$ Ruddlesden-Popper family, in which $n$ 2D layers of $BO_6$ corner-sharing octahedra are joined along the stacking direction and separated by rock-salt AO layers.

The optimally-doped $n=2$ members of this family ($La_{2-2x}B_{1+2x}Mn_2O_7$ where B = Ca or Sr) behave analogously to the $n=\infty$ manganites in the sense that they undergo an insulating- to metallic-like state (I-M) transition coupled to a ferromagnetic transition at temperatures around 120-140 K; besides, they present a large magnetoresistance in this temperature range [2]. However, the change in dimensionality and resulting pronounced cation dependence of the electronic properties can produce physical properties which contrast strongly with the perovskite systems [3-6].

In the $n=2$ member $La_{2-2x}Sr_{1+2x}Mn_2O_7$ an extremely rich variety of magnetic phases have been found as a function of the Sr-doping. One of the most interesting regions extends from $x=0.3$ to $x=0.4$ where ferromagnetic metals (FMM) are found. However, also within a so relatively narrow doping range several different kinds of arrangements of manganese ions spin occur. A common property is the presence of a 2D ferromagnetic ordering in each perovskite layer and between the $MnO_2$ layers even though with different spin directions as a function of $x$. At $x=0.3$, the magnetic moments of each $MnO_2$ layer couple ferromagnetically within a bilayer and antiferromagnetically, along the $c$-axis, between successive bilayers. At $x\approx0.32$ the inter-bilayer coupling becomes FM but still directed along the $c$-axis. At $x\geq0.33$ the magnetic moments direct along the $ab$-plane. The magnetic coupling between the constituents single $MnO_2$ layers changes from FM into canted-antiferromagnetic (AFM) beyond $x\sim0.4$ [7]. Also the magnetic coupling above the transition temperature ($T_C$) is rather interesting. As suggested by several groups [8, 9] a 2D ferromagnetic coupling within the plane should occur and the presence of a FM-AFM correlation between each plane in the bilayer unit, for $x=0.4$, has been pointed out. In any case, there exists two-dimensional ferromagnetic short-range order in a wide temperature region above $T_C$ which may be related to the anisotropic exchange energy ($|J_{ab}|>|J_c|>>|J'|$, $J_{ab}$, $J_c$, and $J'$ standing for the in-plane, inter-single-layer, and inter-bilayer exchange interaction, respectively) in the quasi-two-dimensional FM system.

As in the perovskite manganites, a close coupling between the magnetic and transport properties is observed. However, due to the strong anisotropy that characterizes these layered systems, peculiar features have been observed. Always considering the $0.3\leq x\leq 0.4$ range for $La_{2-2x}Sr_{1+2x}Mn_2O_7$, single-crystals studies have shown that the ratio between the resistivity along the *c*-axis ($\rho_c$) and the resistivity in the *ab*-plane ($\rho_{ab}$) is as large as $10^2$ at room temperature (RT), which suggests a confinement of the carrier motion within the $MnO_2$ bilayer. Usually, the on-set of the long-range FM order is accompanied by a resistivity drop (IM transition) followed by a metallic-like transport. In the *T*-range above $T_C$ both $\rho_c$ and $\rho_{ab}$ show an activated-like transport for $x=0.4$ with hopping energies around 30-40 meV [10]. By lowering the doping the nature of the insulating state in the $\rho_{ab}$ is progressively suppressed and, at $x=0.3$, a metal-like transport ($d\rho_{ab}/dT > 0$) is observed in the range $T_C\leq T\leq 270$ K [2].

The role of A-site doping on the bilayered manganites properties has been studied as done in the case of the perovskites manganites. However, fewer dopants have been considered and for less extensive doping levels due not only to the fact that the field of bilayered manganites is relatively new with respect to that of the perovskites, but also due to less straightforward preparation routes.

Calcium doping on the A-site has been object of an early investigation by Asano *et al.* who studied the $La_{2-2x}Ca_{1+2x}Mn_2O_7$ solid solution for $0\leq x\leq 0.5$ [11]. The main conclusions of that work is the clear evidence of two distinct type of FM ordering which possibly results from anisotropic exchange interactions. Moreover, the material, in the doping region $0.22\leq x\leq 0.5$, undergoes two transitions from a paramagnetic-insulator to a ferromagnetic-insulator and finally to a ferromagnetic- metal with decreasing temperature. A model considering the *intra*- and *inter*-layers spin arrangement was proposed.

Taking into account the available literature regarding the synthesis and characterization of bilayered manganites it appears that further work is needed. In particular we were interested in looking at the role of mixed cation doping on the A-site of the structure. Up to now only one work [12] and our previous investigations [13,14] considered the $(La)_{1.4}(Sr_{1-y}Ca_y)_{1.6}Mn_2O_{7\pm\delta}$ solid solution. However, in Ref. 12 only resistivity measurements are given. We could observe that a partial replacement of Sr with Ca, by keeping the hole doping fixed, is able to change the long range magnetic order from FM to AFM [13], in a quite analogous way as the increasing of the Sr-doping does. This was connected to a size effect induced by the Ca doping. In this paper we further increased the Ca replacement for Sr by reaching the solubility limit of this solid solution, namely $La_{1.4}Sr_{0.8}Ca_{0.8}Mn_2O_7$. In particular, we were interested in looking at the nature of the magnetic structure of this composition and also to the role of oxygen stoichiometry variation on its physical properties. For this purpose, we synthesised and characterized the $La_{1.4}Sr_{0.8}Ca_{0.8}Mn_2O_7$ bilayered manganite (we chose this composition since it represents an *optimally* doping analogous to the $x$=0.3 doping in the $La_{1-x}A^{2+}_xMnO_3$ perovskite manganites) with two different Mn-valence states induced by thermal treatments in reducing (argon) and oxidising (oxygen) environments. Characterization was accomplished by means of x-ray and neutron powder diffraction and SQUID magnetometry.

# Experimental Section

The samples were synthesized by solid state reaction starting from suitable amounts of $La_2O_3$ (Aldrich, 99.999%), $Mn_2O_3$ (Aldrich, 99.999%), $SrCO_3$ (Aldrich, 99.99%) and $CaCO_3$ (Aldrich, 99.99%). Pellets were prepared from well mixed powders and sintered at 1223 K for at least four days, during which time they were re-ground and re-pelletized at least twice. The as prepared polycrystalline samples were annealed at 1123 K in pure argon ($p(O_2)=1 \cdot 10^{-6}$ atm) and pure oxygen ($p(O_2)=1$ atm) to obtain a first sample with an average Mn-valence state of about 3.27 and a second sample with an average Mn-valence state of about 3.38 [14].

X-ray powder diffraction patterns (XRPD) were acquired on a Bruker D8 Advance diffractometer equipped with a Cu anticathode. Electron microprobe analysis (EMPA) measurements were carried out using an ARL SEMQ scanning electron microscope, performing at least 10 measurements in different regions of each sample. According to EMPA the sample was found to be highly homogeneous in terms of chemical composition, with compositions close to the nominal ones.

Static magnetization was measured at 100 Oe from 320 K down to 2 K with a SQUID magnetometer (Quantum Design).

Neutron powder diffraction (NPD) data were acquired on the HPRT instrument [15] of Swiss spallation source SINQ at Paul Scherrer Institute (Villigen, CH) with a wavelengths of 1.494 Å and 1.886 Å. Neutron data were refined by means of the FULLPROFILE software [16].

# Results and Discussion

*Room Temperature Neutron Patterns*

Figure 1 shows the refined pattern at 290 K (at λ =1.494 Å) for the $La_{1.4}Sr_{0.8}Ca_{0.8}Mn_2O_7$ bilayered manganite treated in pure argon. The refinement was accomplished with a two phases model where, beside the bilayered manganite, a second impurity phase of perovskite manganite was introduced. The amount of the latter was estimated to be around 3.5(1)%. The structural details of the main phase are reported in Table 1 for both the argon and oxygen annealing treatments.

The unit cell shrinks by about 0.9% as a consequence of the oxidising environment provided by the oxygen annealing. The cell reduction is strongly anisotropic, with the in-plane parameters reducing only slightly (0.07%) while the out-of-plane parameter reducing of about one order of magnitude more (0.77%). The main effect of the thermal treatment in pure oxygen is the oxidation of the Mn ions. The average Mn-valence state of the sample changes from about 3.27 (argon) to 3.38 (oxygen) as we previously determined by means of X-ray absorption spectroscopy at the Mn-K edge [14].

The lattice parameters change reflects the variation in the bond lengths. As can be seen from Table 1, the Mn-O(2) bond length (i.e. the apical bond between Mn and La) seems to be the most sensitive to the hole doping induced by the oxidising treatment and reduces significantly by increasing the Mn-valence state. The Mn-O(1) bond (the apical bond between two octahedra) reduces less and the Mn-O(3) bond (the equatorial bond) is practically unchanged by the oxidising treatment. Overall, it is the Mn-O(2) shortening which determines the anisotropic behaviour of the lattice constants.

This trend in the bond lengths is related to the filling of the $e_g$ band by holes when the Mn-valence state increases. In addition, it has been pointed out previously that the behaviour in the bond lengths reflects a preferred orbital state occupancy of the $e_g$ electrons, in particular there is a shift of electronic density towards the $x^2$-$y^2$ states as the doping increases [17]. This trend, suggested in [17] as a function of the Sr-doping, in this case it is observed by keeping the cation doping fixed and by changing the oxygen content of the samples.

By refining the RT neutron diffraction patterns we tried to look at the possible distribution of Ca and Sr dopants between the two available sites of the $n=2$ Ruddlesden-Popper manganite, *i.e.* in the rock-salt layer (4e: 0 0 $z$) and in the perovskite block (2b: 0 0 ½) . In order to be able to refine the Ca/Sr ratio we kept fixed the occupancy and distribution of the La ions to the one previously determined by Battle and co-workers [18], namely a roughly equal distribution of the lanthanide between the two sites.

The result of the occupancy refinement is reported in Table 1, as well. It has been found that the Sr ions show a marked preference to be located on the La(1) site which, in fact, is filled with La and Sr and a negligible amount of Ca. The remaining Sr and all the Ca-ions are located on the La(2) site. We found a perfect agreement of the occupancies in the two samples (argon and oxygen annealed) suggesting that the thermal treatments do not influence the Sr/Ca distribution. Also the quantitative agreement between the nominal and calculated formula for both samples is very good: $La_{1.4}Ca_{0.85(6)}Sr_{0.77(6)}Mn_2O_7$ for the reduced sample and $La_{1.4}Ca_{0.83(2)}Sr_{0.80(2)}Mn_2O_7$ for the oxidised sample.

The distribution of the two cations may be explained on the basis of their ionic radii and the coordination of the two available A-sites. The perovskite A-site has a coordination number of 12 and La-O bond lengths of 4×2.73085(4) and 8×2.7117(3) – for the oxidised sample, chosen as example; while the A-site in the rock-salt layer has a coordination number of 9 and La-O bond lengths of 1×2.373(5), 4×2.7463(5) and 4×2.586(2). As expected, the larger cation, Sr, preferentially occupies the perovskite layer.

*Low Temperature Neutron Patterns*

Neutron diffraction patterns have been acquired on both samples for temperatures lower than RT with the primary aim of observing the magnetic structure of the samples. Figure 2 shows the 15 K pattern for the argon annealed $La_{1.4}Sr_{0.8}Ca_{0.8}Mn_2O_7$ (at $\lambda$ =1.886 Å) in a reduced region of the pattern where magnetic scattering may be present. As can be seen, there is no trace of new magnetic peaks and the intensity of the structural ones are still well described by only considering the nuclear contribution. This implies that the sample does not possess any long-range magnetic order (LRO). An analogous behaviour is found for the oxygen annealed $La_{1.4}Sr_{0.8}Ca_{0.8}Mn_2O_7$.

Figure 3 shows the molar susceptibility ($\chi_{mol}$) *vs. T* curves at 100 Oe for the two samples. It can be noticed that in both cases there is a progressive rise of the $\chi_{mol}$ by reducing *T*. In addition, for the argon annealed sample, a step rise of the $\chi_{mol}$ is found below ~90 K. This slow enhancement of the $\chi_{mol}$ suggests that in the samples a magnetic interaction is developing between the Mn ions without, however, a defined transition temperature. We remark that these two samples present an IM transition at about 100 K (reduced sample) and 130 K (oxidised sample) [14].

The structural parameters of the two samples at low *T* may give some insight about the system evolution with *T*. Cell volumes and *c*-parameter for the two samples are reported in Figure 4 for some temperature values below RT. The trends are those commonly observed when a material is cooled down, *i.e.* a progressive reduction of the cell size. A possible anomaly may be found for the volume and *c*-parameter of the argon annealed $La_{1.4}Sr_{0.8}Ca_{0.8}Mn_2O_7$ where, at the lowest measured point (15 K), a small rise of these quantities is observed. However, for the limited set of temperatures considered here we can state that the *c* and *V* parameters do not show any significative discontinuity with *T* (this may be better view through the dotted guides-for-the-eyes-only lines present in Figure 4).

The change of the lattice parameter *a* with temperature is extremely small compared to that of the *c* constant (see Figure 5). This was also previously observed on the same but Ca-undoped bilayered manganite [11]. This small variation of the in-plane constants was attributed to a magnetostriction effect due to the fact that the *a*-axis essentially tracks the development of magnetic order in the *ab* plane. In addition, it may be seen that below 50 K the *a*-axis of the two samples show a deviation. For the reduced sample, after a sort of plateau the *a*-axis contracts while for the oxidised sample, after a slight decrease, the parameter expands below 50K.

The lattice effects shown up to here suggest that in these samples a magnetic interactions within the perovskite planes is present. This can be further confirmed by the behaviour of the Mn-O bond lengths. In Figure 6 and 7 are shown the trends *vs. T* for the equatorial Mn-O(3) and apical Mn-O(2) bonds superimposed to the susceptibility curves. In both cases it can be noticed a tendency of the apical Mn-O(2) bond to expand and of the equatorial Mn-O(3) bond to contract below a certain temperature, which is higher for the reduced sample. However, our set of temperatures does not allow us to define the exact temperature at which the bond lengths change their behaviour. For sure, we can state that for both samples the Mn-O(2) length at 15 K is higher than that at RT which is a clear sign of a charge transfer between in-plane and apical Mn-O bonds which has been in turn connected to the set up of magnetic ordering in these system.

It may be noted that the behaviour of the *a* parameter and of the in-plane Mn-O bond are different for the two samples below 50 K. A difference is also found in the octahedron distortion parameter, *D* (defined as $D = <\text{Mn-O}_{apical}>/\text{Mn-O}_{equatorial}$), which has a higher value at 15 K than RT for the argon annealed sample (1.028(3)-15 K and 1.024(3)-RT) and the opposite for the oxygen annealed sample (1.010(3)-15 K and 1.014(3)-RT). This fact, even though based on relatively few data points, is strongly indicative that in the two samples the charge carriers at low-*T* tend to (re)distribute between the $e_g$ orbitals, and particularly in the $x^2$-$y^2$ states for the oxygen and in the $3z^2$-$r^2$ for the argon annealed samples. This is in perfect agreement with the results found in this layered manganite as a function of Sr-dopant (*x*) [17].

As already pointed out,the lattice effects and susceptibility data clearly demonstrate that magnetic interactions within the perovskites layers are present in these samples. However, the $La_{1.4}Sr_{0.8}Ca_{0.8}Mn_2O_7$ bilayered manganite, for both treatments, does not show any evidence of 3D FM or AFM order, *i.e.* between successive bilayers (*intra*-bilayers). Nevertheless, a two dimensional FM spin ordering is probably present already at high temperatures, and the first sign of that is the "opening" of the Mn-O(2) and Mn-O(3) bond lengths between 290 and 200 K for the reduced sample and between 200 and 75 K for the oxidised sample [11, 19, 20]. This is in agreement with the picture proposed by Asano [11] which suggested that already at relatively high temperatures (well above the Curie temperature and IM transition) FM interaction within the perovskite layers are present. However, differently with this picture, our samples never develop a 3D LRO, neither at the lowest temperature measured. Based on the presented data, the only magnetic interaction found in the $La_{1.4}Sr_{0.8}Ca_{0.8}Mn_2O_7$ bilayered manganite is related to a short-range 2D FM order. Finally, we stress that, even though 3D ordering is not observed, clear IM transitions are found in these samples [14].

The origin of this peculiar behaviour for the $La_{1.4}Sr_{0.8}Ca_{0.8}Mn_2O_7$ manganite is probably due to a size effect but more probably to the cation size variance. In fact, if we consider the same divalent doping and Mn valence state and look at the $T_C$ of the Sr and Ca doped $La_{1.4}A_{1.6}Mn_2O_7$ (A=Sr and Ca) manganites, values of about 100 K and 85 K are, respectively, found. Moreover, both compositions display well defined long range magnetic structures. So, the pure size effect can not be enough to explain the absence of clear magnetic transition and order in the $La_{1.4}Sr_{0.8}Ca_{0.8}Mn_2O_7$ manganite. However, if we consider the significant cation size mismatch between Ca and Sr we may expect that this could significantly influence the physical properties of these bilayered manganites as previously observed for the perovskite manganites [21]. In addition, it can be noticed that the A cation variance is able to depress the magnetic transition more than the "simple" size effect [21]. Just as a reference, we previously studied another Ca/Sr mixed layered manganite with the same hole doping: $La_{1.4}Sr_{1.2}Ca_{0.4}Mn_2O_7$. In this case we found a $T_C$ of ~33 K [14]. The statistical variance in the ionic radii distribution is, for the same hole doping, 0.0090 $Å^2$ for $La_{1.4}Sr_{0.8}Ca_{0.8}Mn_2O_7$ and 0.0077 for $La_{1.4}Sr_{1.2}Ca_{0.4}Mn_2O_7$. Unfortunately, up to now, we are not able to find in the current literature more examples to enrich this first observation of a possible cation size mismatch effect in layered manganites. Finally, we may not exclude that the peculiar arrangement between the two A-sites of Ca and Sr plays an additional role in suppressing the magnetic three dimensional spin ordering.

## Conclusion

In this work we have shown, for the first time, the absence of magnetic LRO in an optimally doped bilayered manganite where the A-site of the structure is doped with equal proportions of different isovalent cations (Ca and Sr). The system, however, presents defined IM transitions which suggest that the transport properties are not linked to the evolution of LRO and that two dimensional spin ordering in the layers of the perovskite blocks may be sufficient to "assist" the hole hopping.

We have also shown the ability of oxygen content variation in "simulating" the changes induced by the cation doping. The reduced and oxidised samples showed small but significant differences in their lattice effects, with the reduced sample resembling the $x$=0.3 compound of the $La_{2-2x}Sr_{1+2x}Mn_2O_7$ solid solution and the oxidised one showing features more similar to the $x$=0.4 composition.

Possible reason for the suppression of magnetic order induced by the Ca doping is a size effect coupled to the cation size mismatch between the Sr and Ca ions. Further work is planned in order to put in prominence the role of statistical variance in the ionic radii distribution in affecting the physical properties of layered manganites.

# Acknowledgement

Financial support from the Italian Ministry of Scientific Research (MIUR) by PRIN Project (2004) is gratefully acknowledged. LM is grateful to the "Accademia Nazionale dei Lincei" for financial support. Dr. Elena di Tullio is acknowledged for sample preparation. This work was partially performed at the spallation neutron source SINQ, Paul Scherrer Institute, Villigen, Switzerland

**Table 1**

| Site | Structural Parameters | $La_{1.4}Sr_{0.8}Ca_{0.8}Mn_2O_7$ Argon | $La_{1.4}Sr_{0.8}Ca_{0.8}Mn_2O_7$ Oxygen |
|---|---|---|---|
| | $a$/Å | 3.86438(5) | 3.86200(7) |
| | $b$/Å | 3.86438(5) | 3.86200(7) |
| | $c$/Å | 19.9802(4) | 19.8249(5) |
| | $V$/ Å$^3$ | 298.376(8) | 295.70(1) |
| La(1) | Occ. Sr | 0.033(3) | 0.033(1) |
| | Occ. Ca | 0.002(3) | 0.002(1) |
| | $B$ | 0.81(7) | 0.76(6) |
| La(2) | Occ. Sr | 0.015(4) | 0.017(2) |
| | Occ. Ca | 0.051(4) | 0.050(2) |
| | $Z$ | 0.31727(15) | 0.31718(16) |
| | $B$ | 0.80(7) | 1.37(7) |
| Mn | $Z$ | 0.09838(29) | 0.09872(31) |
| | $B$ | 0.75(7) | 0.97(8) |
| O(1) | $B$ | 2.5(1) | 1.8(1) |
| O(2) | $Z$ | 0.1982(2) | 0.1975(2) |
| | $B$ | 1.45(9) | 1.7(1) |
| O(3) | $Z$ | 0.09635(2) | 0.09603(2) |
| | $B$ | 1.29(5) | 1.37(4) |
| | $R_{wp}/\chi$ | 5.88/2.98 | 6.36/3.46 |
| | Mn-O(1) | 1.966(6) | 1.957(6) |
| | Mn-O(2) | 1.994(7) | 1.958(6) |
| | Mn-O(3) | 1.9326(2) | 1.9317(2) |

# Figures Captions

**Figure 1** – Neutron diffraction pattern ($\lambda$=1.494 Å) for the reduced La$_{1.4}$Sr$_{0.8}$Ca$_{0.8}$Mn$_2$O$_7$ sample. Red crosses represent the experimental pattern, the black line is the calculated one, while the black line at the bottom is the difference between them. Bragg peaks appear as vertical green lines for bilayered manganite and as blue lines for the perovskite phase.

**Figure 2** – Details of the neutron diffraction pattern for the reduced La$_{1.4}$Sr$_{0.8}$Ca$_{0.8}$Mn$_2$O$_7$ sample at 15 K. Symbols have the same meaning as Figure 1.

**Figure 3** – Molar susceptibility *vs.* *T* curves at 100 Oe for the argon annealed La$_{1.4}$Sr$_{0.8}$Ca$_{0.8}$Mn$_2$O$_7$ (black squares) and oxygen annealed La$_{1.4}$Sr$_{0.8}$Ca$_{0.8}$Mn$_2$O$_7$ (white stars).

**Figure 4** – *c* parameter (stars) and cell volume (squares) for the argon annealed (blue symbols) and oxygen annealed (red symbols) La$_{1.4}$Sr$_{0.8}$Ca$_{0.8}$Mn$_2$O$_7$.

**Figure 5** – *a* parameter for the argon annealed (blue stars) and oxygen annealed (red stars) La$_{1.4}$Sr$_{0.8}$Ca$_{0.8}$Mn$_2$O$_7$.

**Figure 6** – Mn-O(2) and Mn-O(3) bond lengths as a function of *T* for the argon annealed La$_{1.4}$Sr$_{0.8}$Ca$_{0.8}$Mn$_2$O$_7$ superimposed at the susceptibility curve at 100 Oe.

**Figure 7** – Mn-O(2) and Mn-O(3) bond lengths as a function of *T* for the oxygen annealed La$_{1.4}$Sr$_{0.8}$Ca$_{0.8}$Mn$_2$O$_7$ superimposed at the susceptibility curve at 100 Oe.

# Table Caption

**Table 1** – Rietveld refinement results for the La$_{1.4}$Sr$_{0.8}$Ca$_{0.8}$Mn$_2$O$_7$ annealed in argon and oxygen.

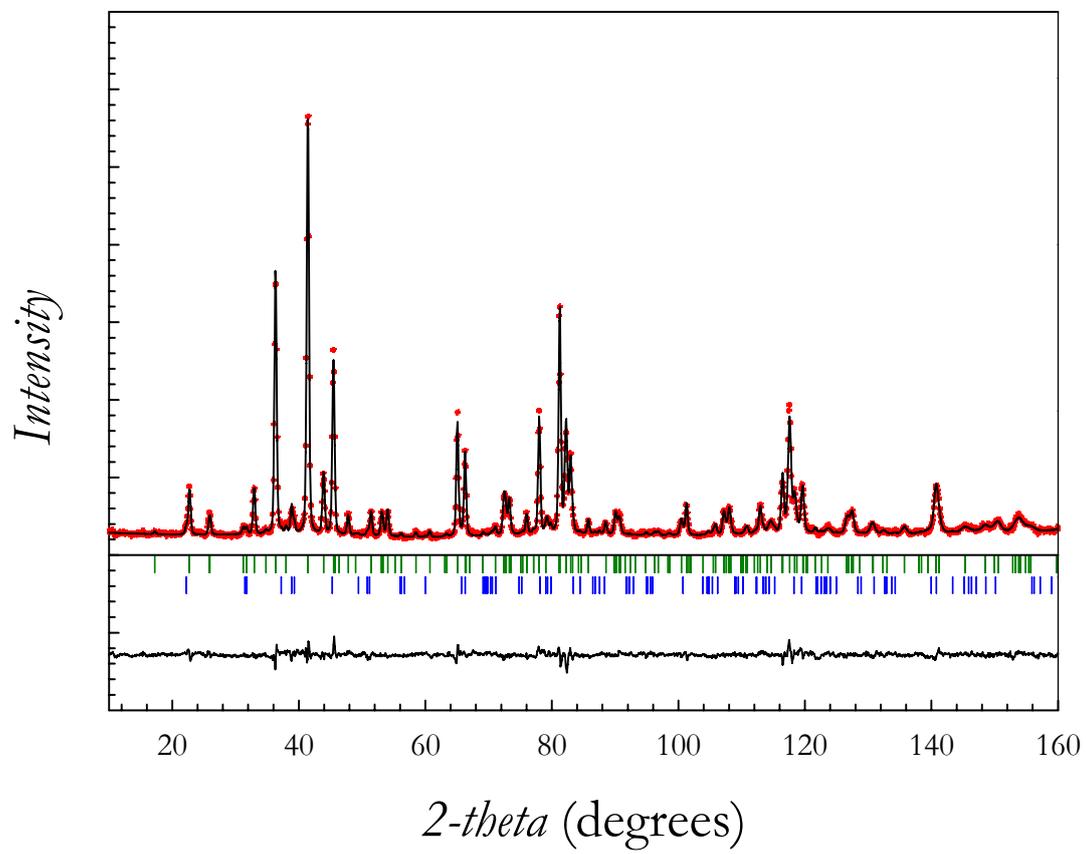

**Figure 1**

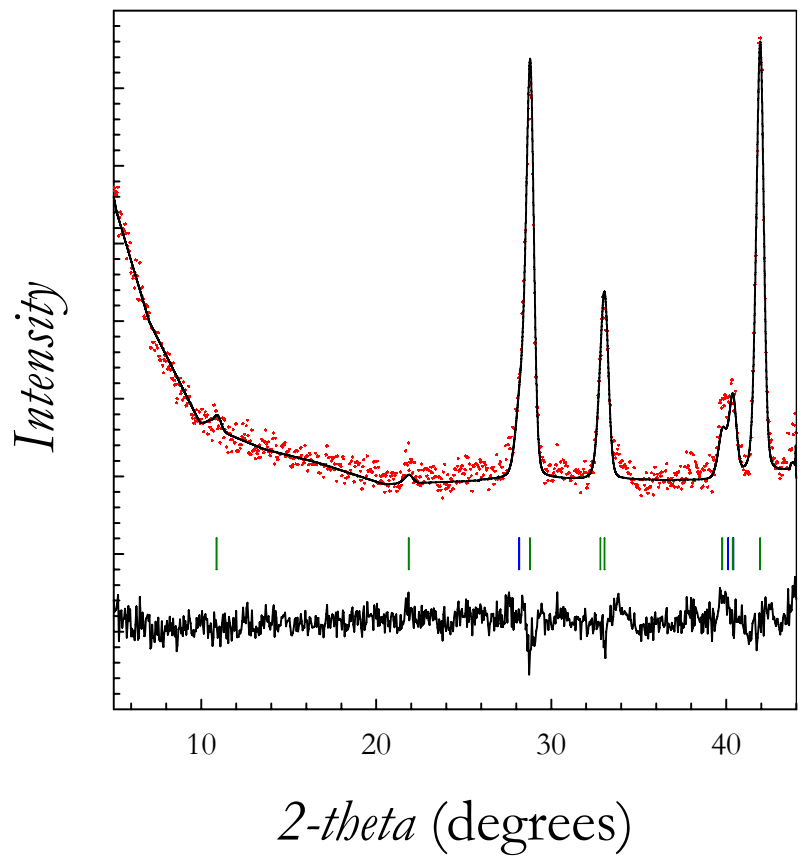

**Figure 2**

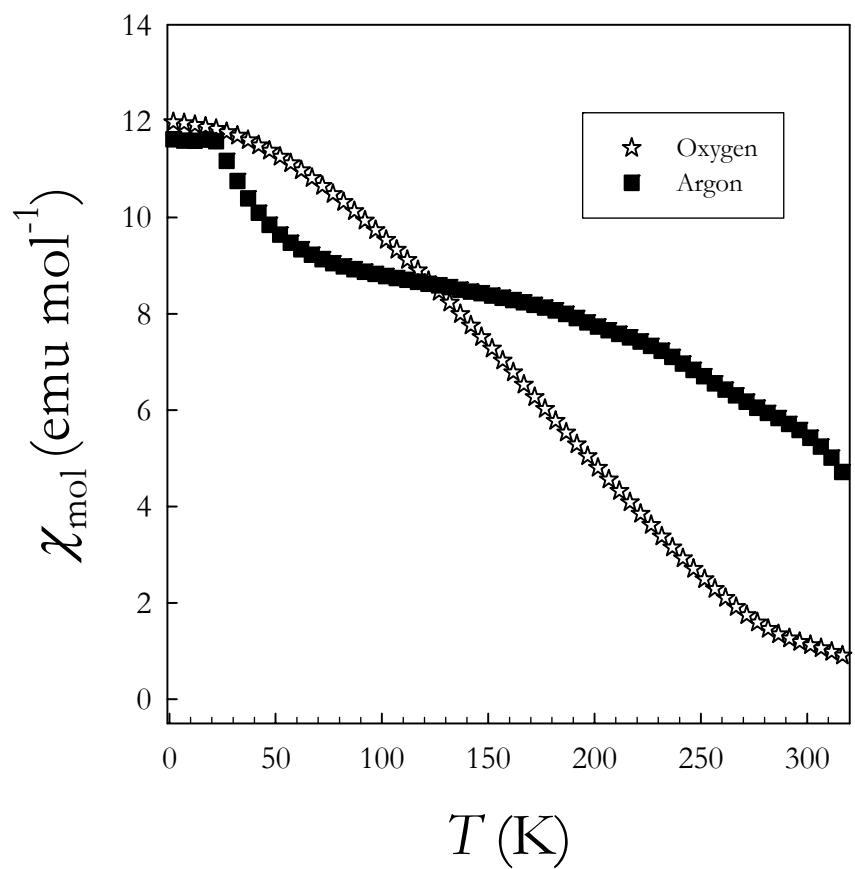

**Figure 3**

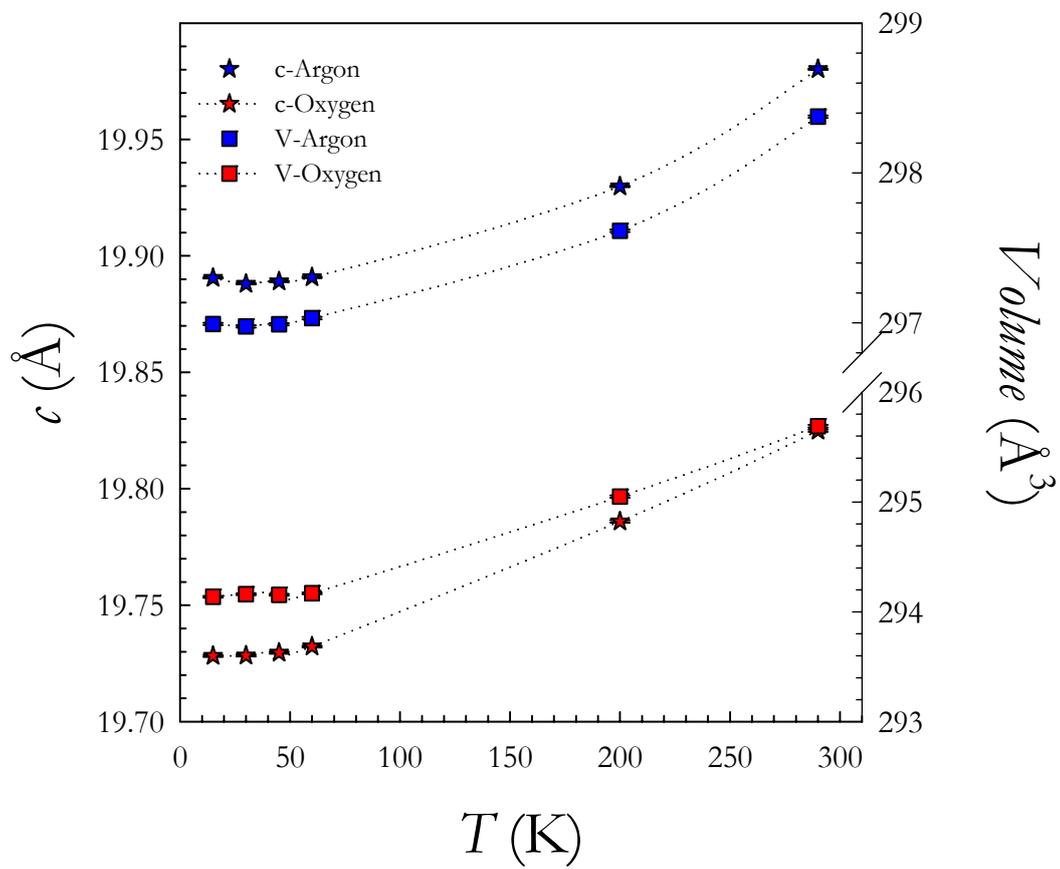

**Figure 4**

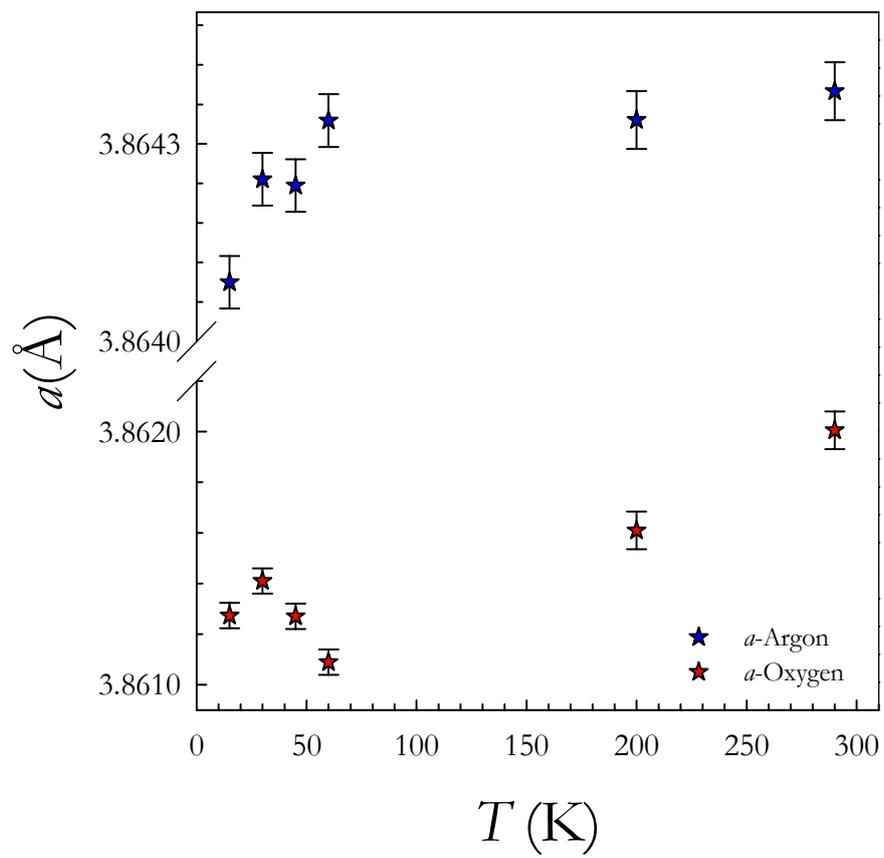

**Figure 5**

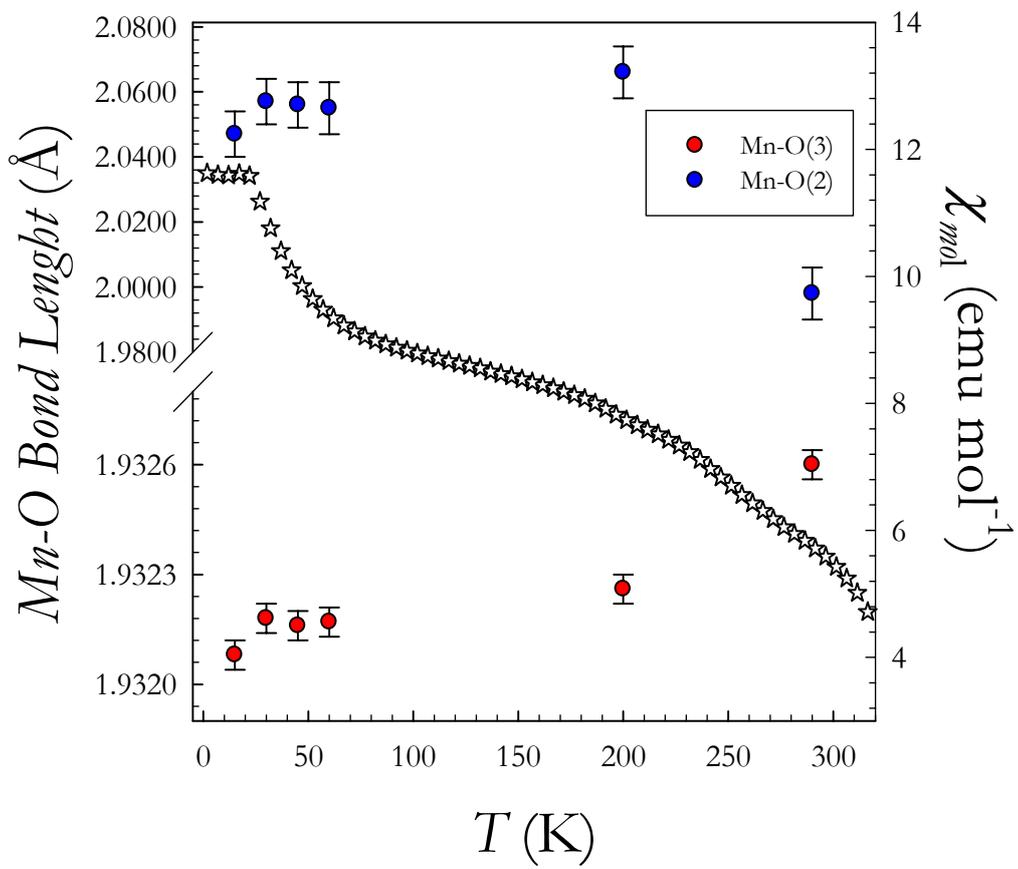

**Figure 6**

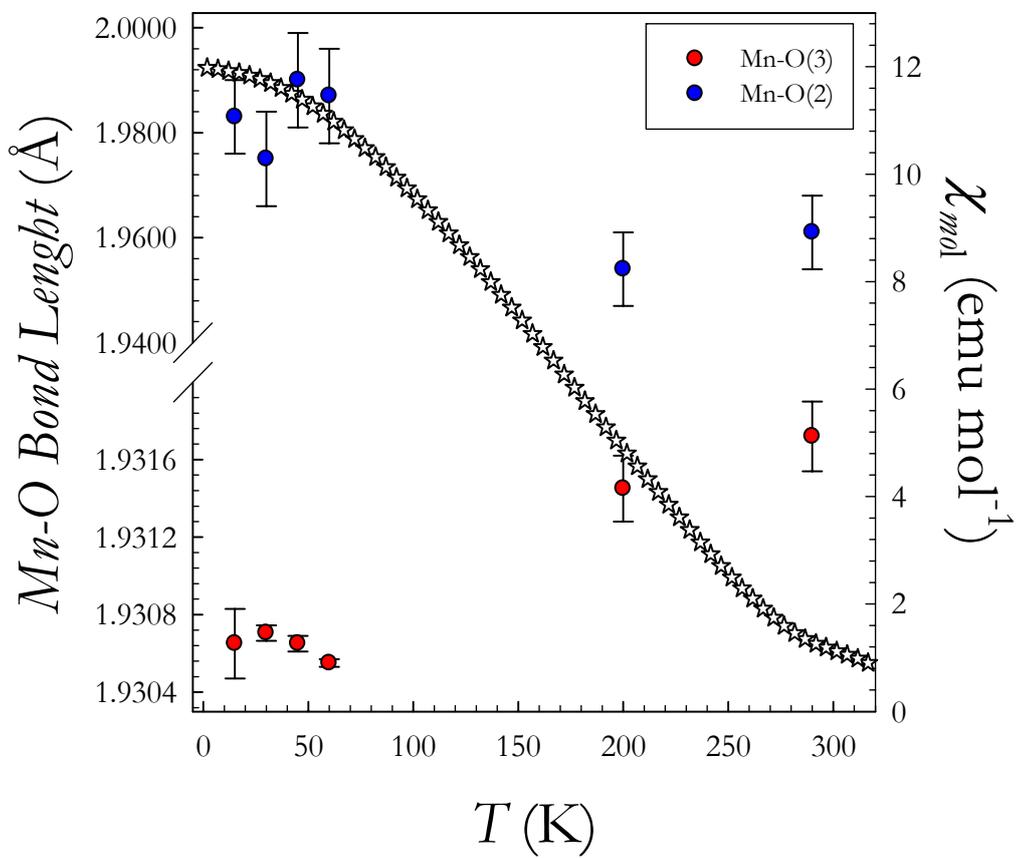

**Figure 7**